\documentclass{aa}

\usepackage{algorithm}
\usepackage[end,]{algpseudocode}
\usepackage{amssymb}
\usepackage{caption}
\usepackage{datetime}
\usepackage{graphicx}
\usepackage{hyperref}
\usepackage{mathtools}
\usepackage{natbib}
\usepackage{placeins}
\usepackage{setspace}
\usepackage{siunitx}
\usepackage{subcaption}

\usepackage{txfonts}
\usepackage{xcolor}

\algnewcommand{\LineComment}[1]{\Statex \(\triangleright\) #1}
\DeclareSIUnit \Jy {Jy}
\DeclareSIUnit \px {px}
\DeclareSIUnit \beam {beam}

\usepackage{xparse}
\makeatletter
\NewDocumentCommand{\LeftComment}{s m}{%
  \Statex \IfBooleanF{#1}{\hspace*{\ALG@thistlm}}\(\triangleright\) #2}
\makeatother
  
\begin{document}

   \title{Simulating realistic radio continuum survey maps with diffusion models}

   % \subtitle{}

   \author{
        T. Vičánek Martínez \inst{1}
        \and H. W. Edler \inst{2}
        \and M. Brüggen \inst{1}
    }
    
   \institute{
        Hamburger Sternwarte, Universität Hamburg, Gojenbergsweg 112, 21029 Hamburg, Germany
        \and
        ASTRON, Netherlands Institute for Radio Astronomy, Oude Hoogeveensedijk 4, 7991 PD Dwingeloo, The Netherlands
   }
   \date{}

  \abstract
  % context heading (optional)
   {The next generation of radio surveys is going to be transformative for cosmology and other aspects of our understanding of astrophysics. Realistic simulations of radio observations are essential for the design and planning of radio surveys. They are employed in the development of methods for tasks, such as data calibration and reduction, automated analysis and statistical studies in cosmology.}
  % aims heading (mandatory)
   {We implemented a software for machine learning-assisted simulations of realistic surveys with the LOFAR telescope, resulting in a synthetic radio sky model and a corresponding artificial telescope observation.}
  % methods heading (mandatory)
   {We employed a diffusion model trained on LoTSS observations to generate individual radio galaxy images with control over the angular size. Single sources are assembled into a radio sky model, using an input catalog from cosmological simulations. We then transformed this sky model into visibilities corresponding to a typical LoTSS pointing. We added realistic noise to this synthetic measurement and obtained our final simulated sky maps through deconvolution. We explored different ways to evaluate our resulting sky model.}
  % results heading (mandatory)
   {We were able to simulate realistic LOFAR observations, covering a sky patch of \ang{5}$\times$\ang{5} at an effective resolution of \ang{;;8.5}. The simulated sources have flux and size distributions that match real observations, and the resulting maps have sensitivities compatible with LoTSS observations. Our diffusion model is able to synthesize high-quality realistic radio galaxy images with precise control over the source sizes. This software can readily be applied to other instruments.}
  % conclusions heading (optional), leave it empty if necessary 
   {}


   \maketitle

%%%%%%%%%%%%%%%%%%%%%%%%%%%%%%%%%%%%%%%%%%%%%%%%
\section{Introduction}
Low-frequency radio surveys play a crucial role in astronomy as they bear a high potential for new discoveries in many areas. These range from the formation and evolution of galaxies and massive black holes to studies of large-scale structure and cosmology \citep{Shimwell+19}. There has been a steady improvement in the depth and resolution of such surveys over the past 60 years. Current and upcoming generations of telescopes are expected to facilitate major breakthroughs in the field \citep{Shimwell+17}. One example of such telescopes is the Low-Frequency Array (LOFAR, \cite{vanHaarlem+13}), with the ongoing LOFAR Two-Metre Sky Survey (LoTSS), aimed at producing a high-resolution survey covering the northern sky at frequencies between 120-168 \si{MHz} \citep{Shimwell+17}.\\
With increased depth and resolution, the massive volumes of data produced by sky surveys pose new challenges in terms of their reduction and processing. This calls for thoughtfully designed instruments, careful planning of observation strategies, and thoroughly optimized calibration procedures. Furthermore, reliable methods for automated analysis of this data are required, incorporating a detailed understanding of its variable characteristics and inherent systematic effects. An important tool that helps deal with these challenges is the use of simulations.\\ 
Throughout the development of a variety of data reduction, calibration, and analysis methods, simulated observational data played a pivotal role. For instance, \cite{Edler+21} implemented the LOFAR Simulation Tool (\textsc{LoSiTo}), a software that simulates LOFAR observations with a realistic modeling of all relevant systematic effects and corruptions. Those simulations were then used to explore novel approaches for an improved calibration of direction-dependent ionospheric effects. Further, \cite{Shimwell+17} used models of radio sources and of the LOFAR high-band antenna (HBA) to assess contamination from bright sources in real observations, the result of which was directly used for calibration of the latter. In another example, \cite{Shimwell+22} simulated LOFAR observations of synthetic sky models to describe source extensions and to quantify the recovery of diffuse emission.\\%
Simulated observations can also play an important part in statistical studies of cosmology. For instance, \cite{Hale+23} investigated the angular two-point correlation function of radio galaxies using data from the second LoTSS Data Release (LoTSS-DR2, \cite{Shimwell+22}). The authors employed the estimator introduced in \cite{LandySzalay93}, which in addition to the real observations requires a sample of random source positions with no spatial correlation. For this, a spatially random catalog was simulated to mimic the source detections expected from the specifics of the LoTSS telescope configuration, accounting for details, such as variations in sensitivity across  the field of view and the survey area.\\
Finally, the development and testing of different kinds of analysis software is likely the most prominent use case for simulated observations. A great example is the second Square Kilometre Array (SKA) science data challenge, whose design and results are described in \cite{Hartley+23}. Participating teams were provided with simulated SKA observations of neutral hydrogen emission, spanning a field of view of 20 square degrees, and complemented by simulated radio continuum observations over the same field. The goal of the challenge was to detect and correctly localize radio sources, and to characterize the different properties of the sources. The project led both to the development of new techniques and to the improvement of existing ones. It provided useful insights for dealing with real observational data in the future; the power of method complementarity was highlighted as one of the main challenge outcomes. This scientific contribution was only made possible through the use of simulated observations, which emphasizes the important role that observations can play in the design and development of next-generation telescopes and surveys and of methods for analyzing the resulting data.\\
In order to serve as a reliable surrogate of real observations, it is crucial that simulations provide an accurate representation of all relevant characteristics present in real data, regarding both the nature of the observed radio signal as well as features introduced by instrumental effects. This includes not only realistic morphologies of single objects, but also accurate distributions of their properties, such as sizes, fluxes, and even morphological classes across different objects. Similarly, the spatial distribution of radio sources should represent that observed in our universe, which is not simply random, but shows some clustering and a dipolar excess (see, e.g., \citealt{Hale+23}). Furthermore, simulated observations should include a precise modeling of the telescope response, accounting for its capabilities in signal recovery, and for noise and artifacts induced by the measurement and data processing. This is relevant not only in the development of applications tailored to the specific properties of the instrument, but is also required to properly take into account any detection and selection biases when simulations are used in statistical cosmology.\\
In this work, we present a software program for simulating realistic LOFAR HBA maps. The generated data mimics observations of a typical LoTSS \SI{8}{\hour} pointing at frequencies in the range 120-168 \si{MHz}, resulting in radio sky maps of \ang{5}$\times$\ang{5} in size, with an effective resolution of \ang{;;8.5}. The simulations are based on a sky model that emulates the signal obtained from sources with realistic properties. This model was created by assembling single-source images, some of which were obtained from a deep generative model that is capable of replicating complex radio galaxy morphologies. The catalog of galaxies injected into the sky model was obtained from the \textsc{t-recs} simulation \citep{Bonaldi+18}, which models the properties of different populations of radio galaxies and derives their spatial distribution from cosmological simulations. The morphologies of extended sources associated with an active galactic nucleus (AGN), which often exhibit complex shapes, are synthesized with a diffusion model trained on LoTSS-DR2 images and conditioned on the source sizes, based on the work presented in \cite{VicanekMartinez24}. Unresolved sources and  star-forming galaxies (SFGs) are modeled as 2D Gaussians. Visibilities corresponding to a LOFAR observation of the constructed sky model are simulated by using \textsc{DDFacet} \citep{Tasse+18} in predict-mode, and realistic measurement noise is added with \textsc{LoSiTo}. The final sky maps are then generated by running the \textsc{DDFacet} implementation of the Högbom-\textsc{clean} algorithm \citep{Hogbom1974} on the simulated visibilities.\\
This paper is structured as follows. In Section~\ref{sec:sub:diffusion-model} we describe our implementation of the generative model we used to synthesize images of extended AGN, covering also the procedure for curating the training data. In Section~\ref{sec:simulation-of-radio-survey-maps} we explain the details of our multi-step approach to the simulation of radio sky survey maps. We introduce our criteria for quality assessment in Section~\ref{sec:quality-evaluation-criteria}, before presenting the results in Section~\ref{sec:results}, and finally discussing our findings in Section~\ref{sec:discussion}.\\
%%%%%%%%%%%%%%%%%%%%%%%%%%%%%%%%%%%%%%%%%%%%%%%%
\section{Diffusion model} 
\label{sec:model-and-training}

Diffusion models (DMs) constitute a class of generative models that have shown great success in image generation over the past years. These type of models make use of a neural network (NN) that is trained to remove artificially added Gaussian noise of different magnitudes from the original training images. Once trained, the NN is then used on a seed image that contains pure Gaussian noise, and called multiple times in an iterative way, at each step removing only a small level of noise. The result of this procedure is a novel image that resembles a sample drawn from the training data distribution, i.e., in the case of this paper a realistic image of a radio galaxy.\\
In the field of natural imaging, DMs have proven to overcome the limitations of other methods employed for image generation, such as variational autoencoders (VAEs) and generative adversarial networks (GANs). VAEs typically show limited capabilities in reproducing fine, high spatial frequency details, resulting in blurry images \citep{Vivenkanthan2024}. While GANs are known to produce high-quality outputs, they often suffer from instabilities during training and are prone to mode collapse, where the network tends to narrow down on a particular part of the training distribution, leading to limited variability in its generated images \citep{Vivenkanthan2024}. DMs are able to combine both image quality and variability, albeit at the cost of increased computational requirements.\\
In past years a few attempts at generative modeling in the context of radio-astronomical imaging have been undertaken. These include the use of VAEs \citep{Bastien+2021}, GANs \citep{Rustige+2023}, and also DMs \citep{Mishra+2024}. Our implementation is based on the work described in \citep{VicanekMartinez24}, which we refer to for a more detailed theoretical elaboration on DMs. The following is a brief summary with focus on aspects that were changed with respect to the previous implementation.

%-----------------------------------------------
\subsection{Training data} 
\label{sec:sub:training-data}

\begin{figure}
    \centering
    \resizebox{\hsize}{!}{\includegraphics{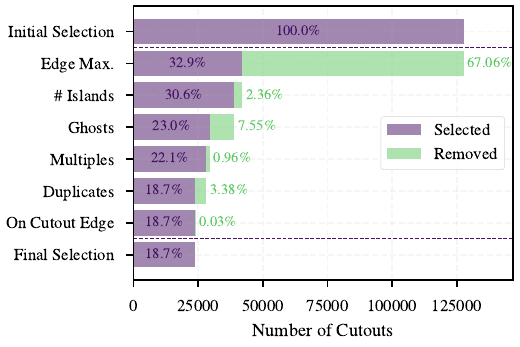}}
    \caption{Bar chart showing the numbers of selected and excluded images at each subsequent step of data selection, as explained in Section~\ref{sec:sub:training-data}. Percentages are given relative to the total number of images in the initial selection.}
    \label{fig:training-data-selection-barchart}
\end{figure}

To create the set of images that the DM is trained on, first a pre-selection of images was obtained from publicly available LoTSS-DR2 data \citep{Shimwell+22}. This selection was then further refined through the application of various quality criteria.\\
We started by retrieving the mosaics of the LoTSS-DR2 Stokes I continuum maps, which have a nominal resolution of \ang{;;6} and are sampled at \ang{;;1.5}/\si{\px}. We created square-shaped cutouts from these mosaics around source positions, which we obtained from the optical cross-match catalog described in \cite{Hardcastle+23}. At the pre-selection stage, we took cutouts with a side length of \ang{;;300} (\SI{200}{px}), centered around the position of the optical counterpart if available in the catalog, otherwise around the position associated with the actual radio source.\\
A quantitative overview of the image selection procedure is given in Fig.~\ref{fig:training-data-selection-barchart} and Table~\ref{tab:selection_cuts}. We first removed all images that contained invalid pixel values. In order to remove images where the source and background signals are not clearly distinguishable, we defined a quantity called $S/N_\sigma$ that serves as a proxy for the signal-to-noise ratio (S/N). For this, we divide the image into a source region and a background region, where the source region is defined by all pixels whose value is higher than five times the image's clipped standard deviation, calculated using the sigma\_clipped\_stats method from the \textsc{astropy} package \citep{astropy22}. For images where no pixel meets this criterion, the multiplier is iteratively reduced from five in decrements of 0.5 until a source region is found. We then calculate the respective mean values in both regions, $\bar{x}_\mathrm{src}$ for the source region and $\bar{x}_\mathrm{bg}$ for the background region, and the $S/N_\sigma$ is finally calculated as the ratio of the two, reading
\begin{equation}
    S/N_\sigma = \frac{\bar{x}_\mathrm{src}}{\bar{x}_\mathrm{bg}}. \label{eq:sigma-SNR}
\end{equation}
We computes this value for every cutout individually and selected only those with $S/N_\sigma \geq 5$. This set of cutouts is referred to as the initial selection in Fig.~\ref{fig:training-data-selection-barchart}.\\
For the selected images, we employed a more sophisticated method of identifying regions on the images corresponding to source emission, which in the following are referred to as source masks. We did so by using islands identified with the Python Blob Detector and Source Finder (\textsc{PyBDSF}, \cite{pyBDSF}), and further refining them with a series of post-processing steps described in Appendix \ref{app:determination-of-source-masks}.\\
An accurate determination of the source masks in the training images provides us with three benefits. First, we can characterize properties based on pixel values that can be attributed to the source emission separately from the background signal, allowing  a more targeted definition of data selection criteria. Second, we can entirely remove the background signal from the selected training images, allowing the generative model to learn and later sample only the features relevant for the simulations, which are the signals associated with clearly detected, extended radio sources. Third, we now have control over placement of the sources on the training images, which allows us to ensure that the source is entirely inside of the inner circle of the square-shaped training image. This in turn enables us to use random rotations by an arbitrary angle during training, as opposed to only multiples of \ang{90}, without parts of the source being cut off. Using this method of data augmentation significantly increases the variety of training samples and thereby provides a denser sampling of the training data manifold.\\
We continued our data selection procedure by discarding all source mask islands that have a pixel value on the edge of the island exceeding 0.05 times the maximum value within the mask, which constitutes the most restrictive step of data selection. This step ensures that we only select sources that are clearly distinguishable from the background and prevents us from including examples where significant parts of the emission are cut away, which results in hard edges that are not representative of real source morphologies. We continued by excluding all images that have more than a single source mask island. Further, by comparing the source masks with the coordinates in the catalog, we were able to exclude cases where source masks do not overlap with any source coordinate (ghosts), or where they overlap with more than one coordinate (multiples). We also excluded any duplicate images (duplicates), i.e., cases where the same source is recognized on different training images, by keeping the ones where the respective source is closest to the image center. Finally, we excluded all images where the source mask overlaps with the outer edges of the image, indicating that the source is only partly contained on the image. \\
A few examples of excluded images, as well as examples of cutouts present at different stages of the selection procedure, are shown in Appendix \ref{app:example-cutouts}. We processed the selected images by applying the identified source mask, i.e., setting all pixels outside the mask to zero. Pixel values were then rescaled such that the minimum pixel is zero and the maximum pixel value is 1. Further, we calculated the smallest outer circle that fully contains the source mask, and relocated the source on the training image in such a way that the center point of the circle coincides with the center point of the image.\\
The result of our selection procedure is a set of $\num{23817}$ training images that each contain a single source and are free of background signals. By ensuring that the source signal is entirely contained in the inner circle of the square-shaped image, all training samples can be freely rotated by arbitrary angles as a method of data augmentation, without losing any information by cutting away parts of the source.\\
Albeit beneficial for different reasons, it is important to understand any possible biases and selection effects induced by this pre-processing and selection procedure, as those directly propagate into sets of images generated by the DM. This aspect is addressed in Sections \ref{sec:sub:diffusion-model} and \ref{sec:discussion}.\\

\begin{table}
    \caption{Numbers of selected and excluded images at each  step of data selection.}
    \label{tab:selection_cuts}
    \centering
    \begin{tabular}{l l l}
        \hline\hline
        Step & Removed & Selected \\
        \hline
        Initial Selection & \num{0} & \num{127559} \\
        Edge Max. & \num{85538} & \num{42021} \\
        \# Islands & \num{3007} & \num{39014} \\
        Ghosts & \num{9632} & \num{29382} \\
        Multiples & \num{1220} & \num{28162} \\
        Duplicates & \num{4306} & \num{23856} \\
        On Cutout Edge & \num{39} & \num{23817} \\
        \hline
        Final Selection & \num{0} & \num{23817} \\
        \hline
    \end{tabular}
    \tablefoot{Selection steps as explained in Section~\ref{sec:sub:training-data}. These are the same numbers illustrated in Fig.~\ref{fig:training-data-selection-barchart}.}
\end{table}

\begin{figure}
    \centering
    \resizebox{\hsize}{!}{\includegraphics{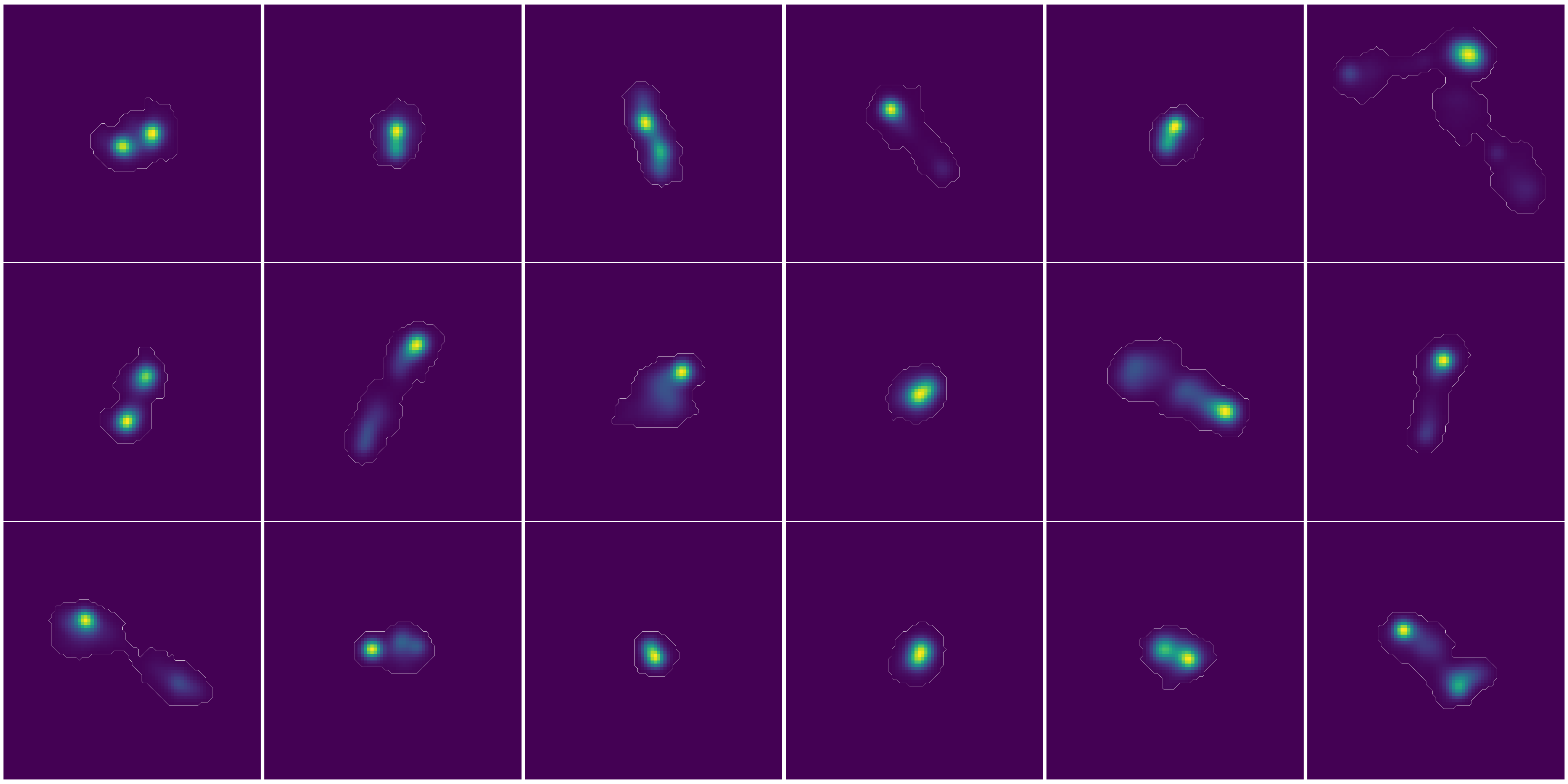}}
    \caption{Random examples included in the final training dataset, selected and processed as explained in Section~\ref{sec:sub:training-data}. The white contours indicate the borders of the source masks.}
    \label{fig:image-example-grid-prototypes}
\end{figure}

%-----------------------------------------------
\subsection{Model architecture and training}
\label{sec:sub:model-architecture-and-training}

The network architecture and training procedure are, with the exception of data augmentation, identical to the ones used in \cite{VicanekMartinez24}, using a U-Net \citep{Ronneberger+15} as a network and employing training with continuous time steps on images of \SI{80}{\px} side length. The only change is that we employed random rotations of an arbitrary angle, rather than multiples of \ang{90}. We conditioned the network on the source size, which we parameterized as the maximum distance between two pixels on the source mask, measured in pixel width. This allows us to control the size during sampling, which is required for creating sky maps with realistic distributions of source sizes. We scaled the size values with a Box-Cox transform \citep{BoxCox64} before being passed to the network, such that their distribution over the training dataset is close to zero mean and unit variance. We excluded 571 images with source sizes larger than \SI{80}{\px}, leaving \num{23246} training samples, which we reduced to the desired size by center-cropping the \SI{200}{\px} images from the original dataset. A random selection is shown in Fig.~\ref{fig:image-example-grid-prototypes}. Before passing the images to the model, we rescaled the pixel values from $[0, 1]$ to $[-1, 1]$. We trained the model for \num{150000} iterations, otherwise the hyperparameters were the same as for the LOFAR model in \cite{VicanekMartinez24}. To improve reproducibility, we include more details on the training and model architecture in Appendix \ref{app:details-on-model-architecture-and-training}.
%%%%%%%%%%%%%%%%%%%%%%%%%%%%%%%%%%%%%%%%%%%%%%%%
\section{Simulation of radio survey maps}
\label{sec:simulation-of-radio-survey-maps}

\begin{figure}
    \centering
    \resizebox{0.9\hsize}{!}{\includegraphics[trim=7mm 8mm 7mm 8mm, clip]{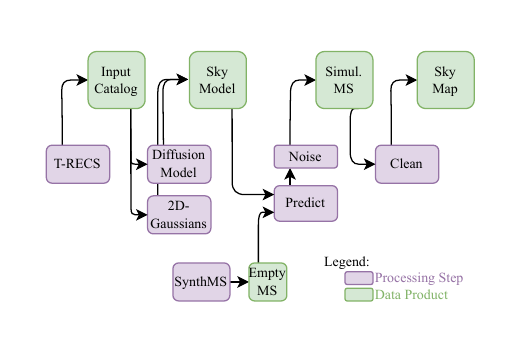}}
    \caption{Overview of different steps and data products of the simulation framework.}
    \label{fig:simulation-steps}
\end{figure}

Our simulated survey maps were generated in a multi-step approach, a schematic overview is presented in Fig.~\ref{fig:simulation-steps}. We started by running the \textsc{t-recs} simulation to generate realistic source catalogs. Based on those, we individually modeled the signals of the sources either by sampling from the DM for large AGN-like sources, or as a 2D Gaussian in the case of star-forming galaxies (SFGs) and unresolved AGN. We then assembled the individual source models by adding their signal onto an initially empty array, which represents the simulated sky, at the position corresponding to the source coordinates. This results in what in this work we refer to as the "sky model". We then used this sky model to simulate LOFAR visibilities that mimic a LoTSS observation by running \textsc{DDFacet} in predict mode on a template measurement set, synthesized using \textsc{LoSiTo}'s "synthms" script. The \textsc{LoSiTo} "noise" step was subsequently applied to add realistic noise to the simulated measurement. Finally, the simulated sky map was obtained by using the Högbom-\textsc{clean} algorithm implemented in \textsc{DDFacet}. In the following, we give a detailed description of the individual steps.

%-----------------------------------------------
\subsection{Input catalog}
\label{sec:sub:input-catalogue}

We ran the \textsc{t-recs} simulation for a \ang{5}$\times$\ang{5} field of view (FOV), which is the maximum size supported for the simulation. We set the output catalog to contain integrated flux density values at a single frequency of \SI{144}{\MHz}, which corresponds to the central frequency of the LoTSS-DR2 Stokes I continuum maps. We set the lower flux limit to \SI{0.08}{\milli \Jy} at the simulated frequency, coincident with the lower limit of the LoTSS catalog. All other parameters were left to their default values.\\
The \textsc{t-recs} catalog contains sources belonging to two classes, SFGs and AGN, each with three different sub-classes. SFGs are divided into late-type, spheroidal and lensed spheroidal SFGs, which differ in their evolutionary model and and dust emission properties. AGN are divided into steep-spectrum sources (SS-AGN), flat-spectrum radio quasars (FSRQs) and BL Lacs. Apart from different spectral and evolutionary characteristics, FSRQs and BL Lacs represent AGN with a small angle between the jet axis and the line of sight, which therefore appear as compact sources, whereas SS-AGN represent sources at a larger inclination with extended morphologies. The output catalog contains several different parameters for every source, out of which we use the source coordinates, integrated flux and angular size for our simulation of the sky model, whereas the latter is only used for sources modeled as 2D Gaussians. While AGN have a single parameter describing the size, defined within \textsc{t-recs} as the intrinsic size of the source projected onto the plane of the sky, SFGs instead have two parameters describing the minor and major axis size, where the galaxy is assumed to have an elliptical shape. In order to avoid the presence of exceptionally bright sources that cause large imaging artifacts in the final product, we excluded all sources that exceed an integrated flux density value of \SI{2}{\Jy}. Based on the counts derived from the LoTSS-DR2 maps, on average we expect about fewer than one source above this threshold for a \ang{5}$\times$\ang{5} sky patch.

%-----------------------------------------------
\subsection{Source images and sky model}
\label{sec:sub:source-images-and-sky-model}

We chose to model all SFGs as elliptical 2D Gaussians, irrespective of their sub-class, where the minor and major axis size are set to the respective values in the \textsc{t-recs} catalog. In the case of AGN, all FSRQs and BL Lacs, as well unresolved SS-AGN, were modeled as circular 2D Gaussians, where the width is again determined by the source size in the catalog. Unresolved SS-AGN are defined as the ones smaller than \ang{;;1.5}, which is the size of a single pixel in the LoTSS maps. Gaussians were generated in such a way that the full width at half maximum (FWHM) of their minor or major axis is half the size of the corresponding axis in the catalog. This choice is consistent with the catalog-based estimate of the largest angular size parameter taken in \cite{Hardcastle+23}. Images of resolved SS-AGN were modeled by sampling from the DM. Since the size parameter used as conditioning during training of the DM is defined via the source mask and therefore differs from the way it is defined in the \textsc{t-recs} catalog, we decided to derive those values by resampling them from the size distribution found in the training dataset. This step is justified because in the \textsc{t-recs} simulation, the size values are uncorrelated to any other source parameters like the flux or position. We modeled the size distribution in the training data by creating a 100-bin histogram of the largest linear sizes of the image masks for every image. Sizes were then sampled by choosing from the bins with probability proportional to the corresponding counts, and then drawing a value within the bin edges from a uniform distribution. For cases where the desired size exceeds the maximum of \SI{80}{\px} imposed by the image size of the DM, we first sampled from the DM using this maximum size value and subsequently upscaled the generated image to the desired size by using the rescale function of the \textsc{skimage} package \citep{vanderWalt2014}.\\
In order to construct the sky model by assembling the single source images, we first generated an empty array that represents the entire \ang{5}$\times$\ang{5} sky model at the LoTSS spatial sampling of \ang{;;1.5}/\si[per-mode=symbol]{\px}, which results in a side length of \SI{12000}{\px}. We then added the generated source images to the empty array, such that the source center coincides with the corresponding source position given in the \textsc{t-recs} catalog. In particular, for Gaussian-like sources, the source center corresponds to the image center by design. For images sampled from the DM, we determined the image centroid through moment analysis.\\
Our DM was trained on real LoTSS restored images, and we directly used the DM to create the sky model. As a consequence, the generated samples are limited by the \ang{;;6} angular resolution of LoTSS, meaning that the size distribution in our sky model differs from the intrinsic sizes by a convolution corresponding to the \ang{;;6} synthesized beam. In order to replicate this effect for the Gaussian-modeled sources and hence make the sky model self-consistent, we additionally applied a Gaussian blur to those images, with a kernel FWHM of \SI{4}{\px}, which corresponds to \ang{;;6} at the LoTSS sampling. This way, all source images have the same implicit resolution limit, regardless of the way in which they were produced. As a final step, we scaled the generated images in such a way that the pixel sum corresponds to the integrated flux catalog value in \si{\Jy}, meaning the images have pixel values corresponding to flux in units of \si{\Jy\per\px}.\\
As is manifested in this last step, the approach taken in this work inherently assumes that morphology and brightness of the sources are intrinsically unrelated, as they are generated independently. In this context, morphology does not refer to the size of the source, for which the relation to flux density is accounted for in the \textsc{t-recs} simulation, but the shape and morphological features. Clearly, more distant sources tend to be both smaller and fainter, which at the resolution limit will cause dimmer sources to have a morphology similar to the restoring beam. However, for extended, resolved sources the morphology is constructed independently of brightness, and the DM is trained on scaled images with no information on the intrinsic flux of the sources. The validity of this assumption is further discussed in Section~\ref{sec:discussion}.\\
%
%-----------------------------------------------
\subsection{Simulated observations and imaging}
\label{sec:sub:simulated-observations-and-imaging}

Observational data obtained from radio interferometry is typically stored in a format called "measurement set" (MS). We used \textsc{LoSiTo}'s synthms script to create a group of template MSs that resemble an observation from a single real LoTSS \SI{8}{\hour} pointing. These MSs were then filled with values by running the \textsc{DDFacet} deconvolution software in predict mode, using our simulated sky model as input image. This results in MSs that contain the data that would be measured if the telescope observed the modeled sky. Subsequently, we added realistic instrument and background noise to the MSs by using \textsc{LoSiTo}'s noise step. Finally, we again used \textsc{DDFacet} to apply the Högbom-\textsc{clean} deconvolution algorithm, from which we obtained the final simulated sky map. Further technical details on the MS metadata and the imaging parameters are given in Appendix \ref{app:technical-details-of-visivilities-and-imaging}.\\
Our approach results in an important difference between the simulated sky maps and the real LoTSS observations used throughout this work. For reasons previously described in Section~\ref{sec:sub:source-images-and-sky-model}, the sky model has an implicit resolution limit of \ang{;;6}, which is ultimately imposed by the restoring beam of the LoTSS maps that provide the training data for the DM. However, the telescope response that is simulated upon this sky model in the subsequent step acts as an additional convolution with the same restoring beam of size $\sigma_\mathrm{Beam}=\ang{;;6}$, which effectively further reduces the resolution of the sky map. In particular, the smallest source in our sky model would be a circular 2D Gaussian with the size of the restoring beam, i.e., $\sigma_\mathrm{Min}=\sigma_\mathrm{Beam}=\ang{;;6}$. This size is then extended by the application of the telescope's point spread function, which is approximately described by a convolution with the restoring beam. Thus, since the size of Gaussian distributions is added in quadrature with the size of the Gaussian convolution kernel, and the FWHM is directly proportional to the standard deviation, the resulting size of the smallest possible source $\sigma_\mathrm{Lim}$ in the sky model is given by
\begin{equation}
    \sigma_\mathrm{Lim} = \sqrt{\sigma_\mathrm{Min}^2 + \sigma_\mathrm{Beam}^2} = \sqrt{2} \cdot \sigma_\mathrm{Beam} \approx \ang{;;8.5}. \label{eq:effective-resolution}
\end{equation}
This value can be interpreted as the effective resolution of our simulated sky maps.

%%%%%%%%%%%%%%%%%%%%%%%%%%%%%%%%%%%%%%%%%%%%%%%%
\section{Quality evaluation criteria} 
\label{sec:quality-evaluation-criteria}

\subsection{Diffusion model samples}
\label{sec:sub:diffusion-model-samples}

In order to evaluate the quality of the samples generated with the DM, we calculated several different metrics, each of them for both the training images and for a set of sampled images. We then compared the two resulting histograms over the calculated quantities. The metrics are based on values of pixels that lie within the source masks, as well as on properties of the masks themselves. Source masks for the generated samples were determined in a similar way as for the training images, with the difference that the local background noise on the samples is induced by the DM sampling process, rather than through observational effects like in the training data. This requires a dedicated way of background noise determination, more details are given in Appendix \ref{app:determination-of-source-masks}.\\
To begin with, we compared the pixel value histograms over the entire datasets, i.e. one count corresponding to a single pixel out of all the images, considering only those pixels that fall inside the image's source mask. The rest of the quantities were individually determined for single images. Those include the pixel value mean and standard deviation, which were again calculated considering only the pixels that lie inside the source mask determined for the respective image. This ensures that the analysis is focused only on the source signal, excluding surrounding regions which will not be used in the simulation. Further, we also evaluated the mask size and mask area, which provide information on the size and extension of the source, respectively. The size is defined, in the same manner as for the conditioning of the DM, as the maximum distance between two pixels on the source mask. The area is defined as the number of masked pixels. Both quantities are measured in pixels, however for the size this refers to a length in units of pixel width, whereas for the area it describes the actual number of pixels. To avoid confusion, we refer to the pixel width as \si{px} to indicate length, and consequently employ the unit of \si{px\squared} to indicate area. Moreover, the ratio between the mask area and outer circle area provides information about the shape. For instance, a source with a large outer circle but comparatively small mask area is likely to be very elongated, while a ratio close to unity indicates a compact shape. We refer to this ratio as the compactness, with its value always lying in the range of $(0, 1)$ since the circle by definition encompasses the entire mask.\\
Finally, in order to assess the quality of the size conditioning, i.e., the capacity of the DM to accurately generate sources of desired size, we evaluated the difference between the input value $s_\mathrm{In}$ used for sampling and the actual size $s_\mathrm{Out}$ of the sampled source. We calculated the size offset $\Delta s$ defined as $\Delta s = s_\mathrm{Out} - s_\mathrm{In}$, for every generated sample and plotted the resulting histogram. In this case, we compare input and output values rather than real and generated images, therefore only one histogram was produced.
\subsection{Properties of the simulated sources}
\label{sec:sub:properties-of-the-simulated-sources}

To assess the realism of our simulations, we compared properties between the simulated radio sky maps and real LoTSS observations through the use of source catalogs. For the LoTSS data, we used the publicly available non-redundant source catalog, generated with \textsc{PyBDSF} as described in \cite{Shimwell+22}. To facilitate a meaningful comparison, we ran \textsc{PyBDSF} on our simulated maps with the same settings in order to obtain an equivalent source catalog.
Based on those catalogs, we compared the distributions of integrated and peak fluxes of the sources, as well as their sizes, quantified as the major axis obtained from image moment analysis on the \textsc{PyBDSF} source model.\\
As previously explained in Section~\ref{sec:sub:simulated-observations-and-imaging}, the simulation of the telescope response induces an additional convolution that results in a blurring of the sky maps as compared to the sky model. This should have no effect on the integrated flux of the sources, however it does change, both, the peak flux values and the source extensions. In particular, as the source signal is spread out by this blurring effect, the peak fluxes are reduced and the sizes are increased. As a direct consequence, these differences are expected to arise in the comparison between source properties in the simulated sky maps and in real LoTSS observations. Following the same reasoning given for the effective resolution in Section~\ref{sec:sub:simulated-observations-and-imaging}, for a single Gaussian source of \ang{;;6} in the sky model, the size would be increased by a factor of $\sqrt{2}$ in the sky map. This factor is reduced with increasing sizes (e.g., less than 0.05 for sources of $\sim$\ang{;;19}), and therefore is negligible for sources with larger extensions. Following a similar argument as for the size reduction, considering the amplitude of a normalized Gaussian kernel with a FWHM of $\sigma_\mathrm{Beam}$ is proportional to $\sigma_\mathrm{Beam}^{-2}$, the peak flux of a \ang{;;6} source should decrease by a factor of two. This effect is again reduced with larger source size. However, as will be covered in Section~\ref{sec:sub:source-fluxes-and-sizes}, quantifying the differences induced by this effect is less straightforward in practice, as residuals of the imaging procedure in the final step of the simulation also influence the sizes and fluxes of sources. Nonetheless, this consideration provides a qualitative picture of what is expected when comparing the histograms over the mentioned source properties.

\subsection{Sensitivity of simulated sky maps}
\label{sec:sub:sensitivity-of-simulated-sky-maps}

A common way of quantifying the sensitivity of a survey is by determining the level of residual noise. We did this by calculating the root mean squared (RMS) for the \textsc{DDFacet} residual image, which should lie below the target LoTSS sensitivity \SI{100}{\micro\Jy\per \beam} mentioned in \cite{Shimwell+22} and ideally match the actual sensitivities of the real sky maps reported therein. The result can be regarded as the sensitivity limit of our simulated maps, averaged over the FOV. In order to obtain spatial information, we divided the residual image into smaller, square-shaped tiles and calculated the RMS for each tile individually. This way, we obtain a view of how the RMS varies across the FOV of our simulated observation. The latter was reconstructed from a single simulated LOFAR pointing, and therefore contains the effect of the LOFAR station beam. This station beam describes the telescope sensitivity profile as a function of the angular distance from the pointing center, and this profile is typically characterized as a circular 2D Gaussian profile with a FWHM of \ang{3.96} at the central frequency of \SI{144}{\mega\hertz} \citep{vanHaarlem+13}. This means that the sensitivity decreases toward the edges of the sky map, i.e., sources farther away from the center will have a comparatively lower apparent flux. This effect is corrected for during the imaging procedure by normalizing the apparent flux with a model of the beam profile, correcting for those lower fluxes while thereby amplifying not only the source signals, but also the noise level. Hence, since the residual RMS is a measure of the underlying noise level, its spatial distribution across the FOV is expected to resemble the inverted beam profile, i.e., higher noise levels toward the edges of the map.

%%%%%%%%%%%%%%%%%%%%%%%%%%%%%%%%%%%%%%%%%%%%%%%%
\section{Results} 
\label{sec:results}

%-----------------------------------------------
\subsection{Diffusion model}
\label{sec:sub:diffusion-model}

\begin{figure}
    \centering
    \resizebox{\hsize}{!}{\includegraphics{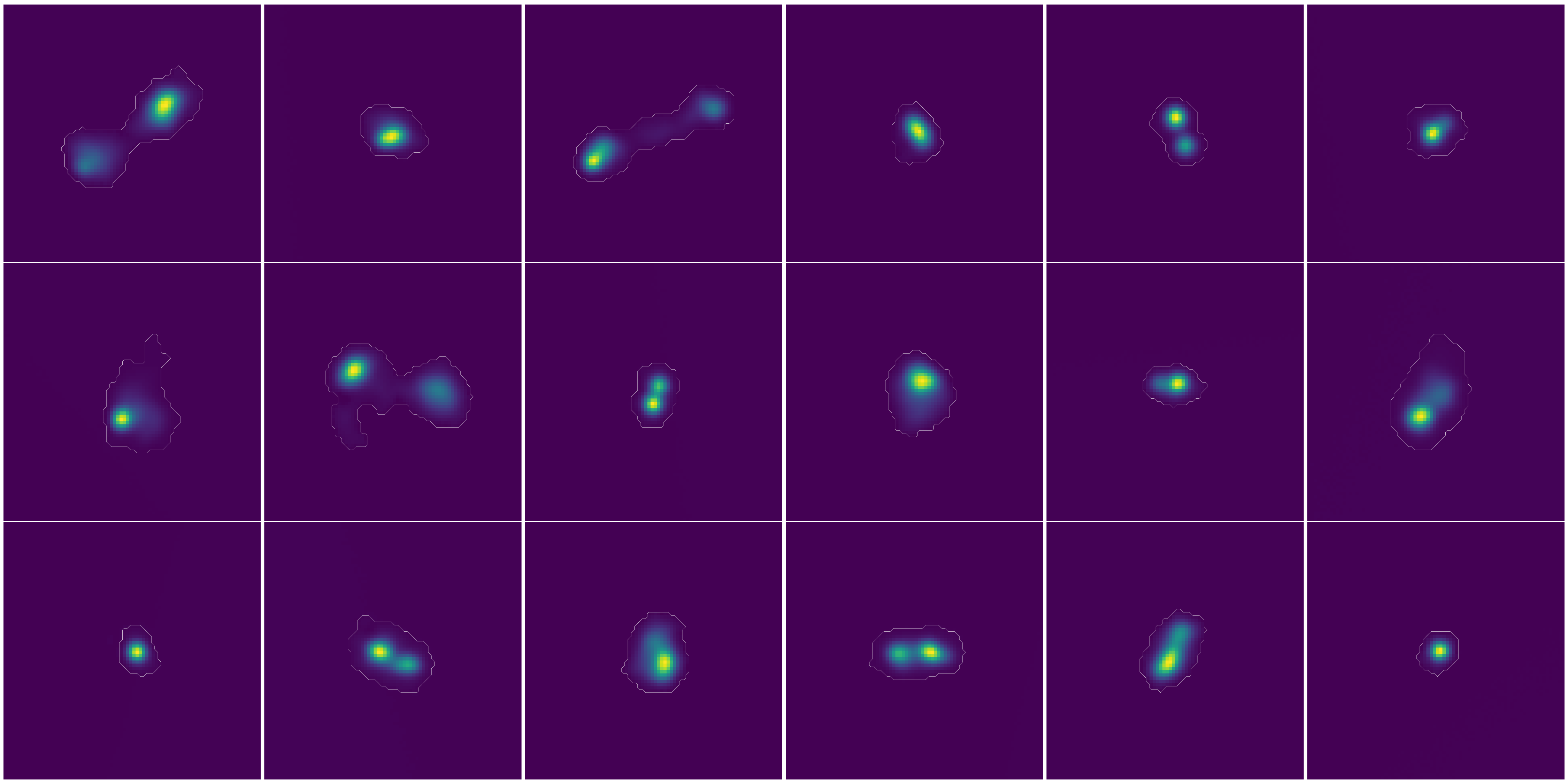}}
    \caption{Random examples of images generated with the DM. The white contours indicate the borders of the source masks}
    \label{fig:image-example-grid-samples}
\end{figure}

\begin{figure*}
    \centering
    \begin{subfigure}[t]{0.45\hsize}
        \centering
        \resizebox{\hsize}{!}{\includegraphics{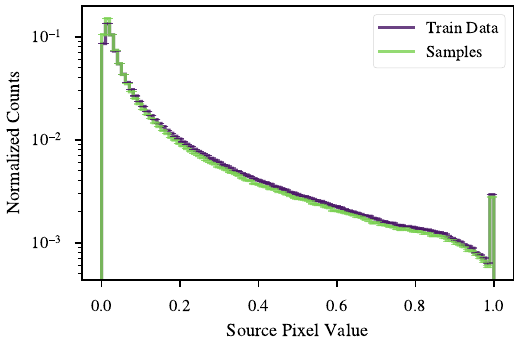}}
        \label{fig:sub:DM-pixel-values}
    \end{subfigure}
    \begin{subfigure}[t]{0.45\hsize}
        \centering
        \resizebox{\hsize}{!}{\includegraphics{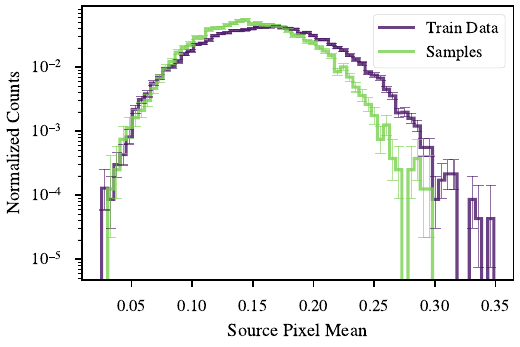}}
        \label{fig:sub:DM-pixel-means}
    \end{subfigure}
    \begin{subfigure}[t]{0.45\hsize}
        \centering
        \resizebox{\hsize}{!}{\includegraphics{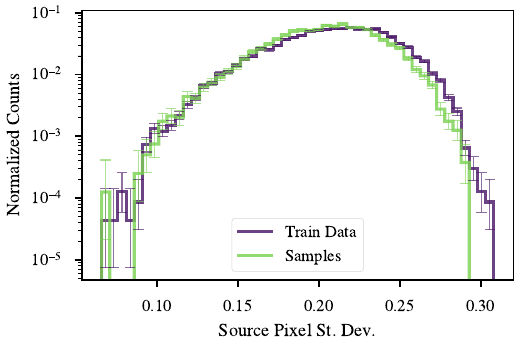}}
        \label{fig:sub:DM-pixel-std}
    \end{subfigure}
    \begin{subfigure}[t]{0.45\hsize}
        \centering
        \resizebox{\hsize}{!}{\includegraphics{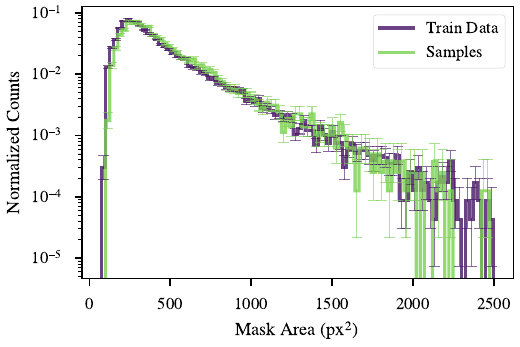}}
        \label{fig:sub:DM-mask-areas}
    \end{subfigure}
    \begin{subfigure}[t]{0.45\hsize}
        \centering
        \resizebox{\hsize}{!}{\includegraphics{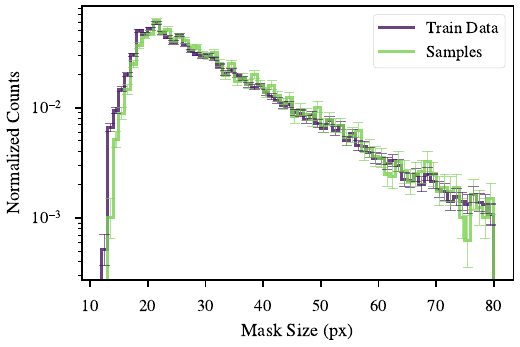}}
        \label{fig:sub:DM-mask-sizes}
    \end{subfigure}
    \begin{subfigure}[t]{0.45\hsize}
        \centering
        \resizebox{\hsize}{!}{\includegraphics{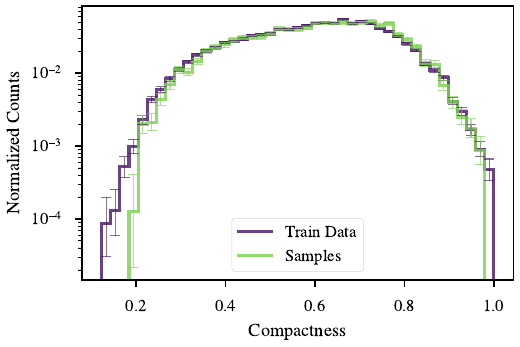}}
        \label{fig:sub:DM-mask-compactness}
    \end{subfigure}
    \caption{Histograms of the metrics introduced in \ref{sec:sub:diffusion-model-samples} showing the distribution over the training dataset and DM-generated images. In the case of pixel values, mean and standard deviation, only pixels that fall within the source mask are included.}
    \label{fig:DM-Histograms}
\end{figure*}

\begin{figure}
    \centering
    \resizebox{\hsize}{!}{\includegraphics{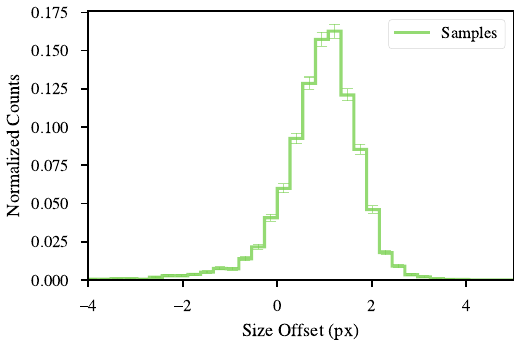}}
    \caption{Distribution of size oﬀsets $\Delta s$ between the size of the source mask identified on the sample and the size passed as input for sampling, as defined in Sect. \ref{sec:sub:diffusion-model-samples}.}
    \label{fig:diffusion-model-size-offsets}
\end{figure}

We evaluated the performance of the DM by comparing properties of images generated with the DM to those of the original training images. For this, we sampled from the DM a set of \num{8000} size-conditioned images after training. Input values for the sizes were drawn from the training data size distribution as described in Section~\ref{sec:sub:source-images-and-sky-model}. Pixel values of the sampled images were rescaled such that the individual maximum value coincides with 1, without considering the minimum value, before determining the source masks. On a single Nvidia A100 GPU, it took \SI{3}{\minute} \SI{16}{\second} to sample a batch of \num{1000} size-conditioned images. A random selection of generated samples and the respective source masks is shown in Fig.~\ref{fig:image-example-grid-samples}. After thorough visual inspection, we found that there are no physically implausible outputs produced by the model.\\
The metrics we evaluated, as they are introduced and explained in Section~\ref{sec:sub:diffusion-model-samples}, are the source pixel values as well as their mean and standard deviation, mask size, mask area and mask compactness. Histograms of the metrics are shown in Fig. \ref{fig:DM-Histograms}. Counts were normalized by dividing through the respective number of images (or pixels in the case of the pixel value distribution in the top left), and error bars correspond to the $1\sigma$ frequentist confidence intervals.\footnote{Calculated using \textsc{astropy}'s stats.poisson\_conf\_interval \citep{astropy22}.}\\
Overall, the distributions found in the training data are replicated remarkably well by the DM samples. In the pixel value histogram on the top left plot, sampled sources seem to have a slightly higher number of pixels close to zero, which in turn induces a small downward shift in the rest of the histogram as compared to the training data. Apart from that, even detailed features like the shape of the histogram between values of 0.8 and 1 are accurately reproduced. We note that the peak at pixel values of 1 is due to the fact that through the scaling procedures, every image in both the datasets has exactly one pixel with a value of 1.\\
The mean value distribution in the top right shows the largest deviation between training data and samples. Both distributions resemble Gaussians, but the sample distribution is slightly skewed toward smaller values. A similar, but visibly smaller trend is observed for the standard deviations in the middle left plot. The opposite is found for the middle right and bottom left histograms, where the sample distributions of size and mask area are slightly skewed toward higher values, i.e., larger sizes. We note that, in the case of the size distribution, the histogram over the training data is the same one that was used to draw sizes as conditioning input for the DM. Hence, we expect those distributions to match if the conditioning mechanism is learned well. The compactness distributions in the bottom right plot show an excellent resemblance over most part of the value range, only deviating at small values below $\sim\!0.15$, where there is a slight excess in the samples distribution, although in a regime of fewer than 10 counts per bin, and therefore probably attributed to shot noise of the sampling process. Finally, the size offset distribution in Fig.~\ref{fig:diffusion-model-size-offsets} is well described in its shape by a Gaussian, with a mean value of \SI{0.89}{\px} and a standard deviation of \SI{0.97}{\px}. Interestingly, the peak occurs at a value of around 1, indicating that the sample masks tend to have a larger size by about one pixel on average as compared to the conditioning input.\\
In order to interpret the differences between real and sampled data, it is important to appreciate that they reflect not just characteristics of the datasets in general, but also the two slightly different methods of how source masks are generated. This is an inevitable consequence of the different nature of the compared image datasets, as the training images are extracted from real observations, where sources are surrounded by diffuse foreground emission and imaging artifacts. In contrast, the sampled images result from training on idealized source images with zero-valued background pixels, hence the background signal on the generated images is purely a residual of the DM's sampling process. The observed discrepancies can be entirely explained by hypothesizing that the sample source masks are generally slightly more inclusive than the ones on the training images. This means that more low-value pixels at the edges of the source are included into the source mask, which explains the slight excess in the pixel value distribution, as well as the shifts toward lower mean values and even lower standard deviations, as pixels around the source are all close to zero and therefore not largely variable. Similarly, this explains the shift in the sample size distribution that is on the order of one pixel, as well as the slight shift toward larger mask areas and trivially the average size offset of one pixel. In fact, we were able to reduce and even invert the observed effects by setting a higher pixel value threshold for defining the sample source masks, which suggests that the procedure of sample source mask generation could potentially be optimized to maximize the similarity in the observed distributions. However, keeping the purpose of the source masks for our simulation in mind, a more inclusive way of defining the sample source masks mitigates the risk of excluding relevant signal at the cost of possibly including very low-level Gaussian noise, which is not expected to make any noticeable difference in the final result. We therefore adhered to the implemented procedure.\\
The only observed discrepancy not explained by the mask sizes is the variance in the size offset observed in Fig.~\ref{fig:diffusion-model-size-offsets}. This seemingly statistical variation is most likely inherent to the conditioned sampling, possibly emerging from the fundamentally statistical nature of the underlying generative model. Further experiments might clarify whether this variance can be reduced by tuning different training hyperparameters or by increasing the training budget. Nonetheless, the variance is reasonably small, as over \SI{99.6}{\percent} of the sources have an offset of magnitude not larger than \SI{4}{\px}, which corresponds to the FWHM of the restoring beam. Hence, we conclude that the sampling control over the source sizes is reliable enough for the purpose of this work.\\

%-----------------------------------------------
\subsection{Source fluxes and sizes}
\label{sec:sub:source-fluxes-and-sizes}

\begin{figure*}
    \centering
    \includegraphics[width=17cm, trim=0 2mm 0 5mm, clip]{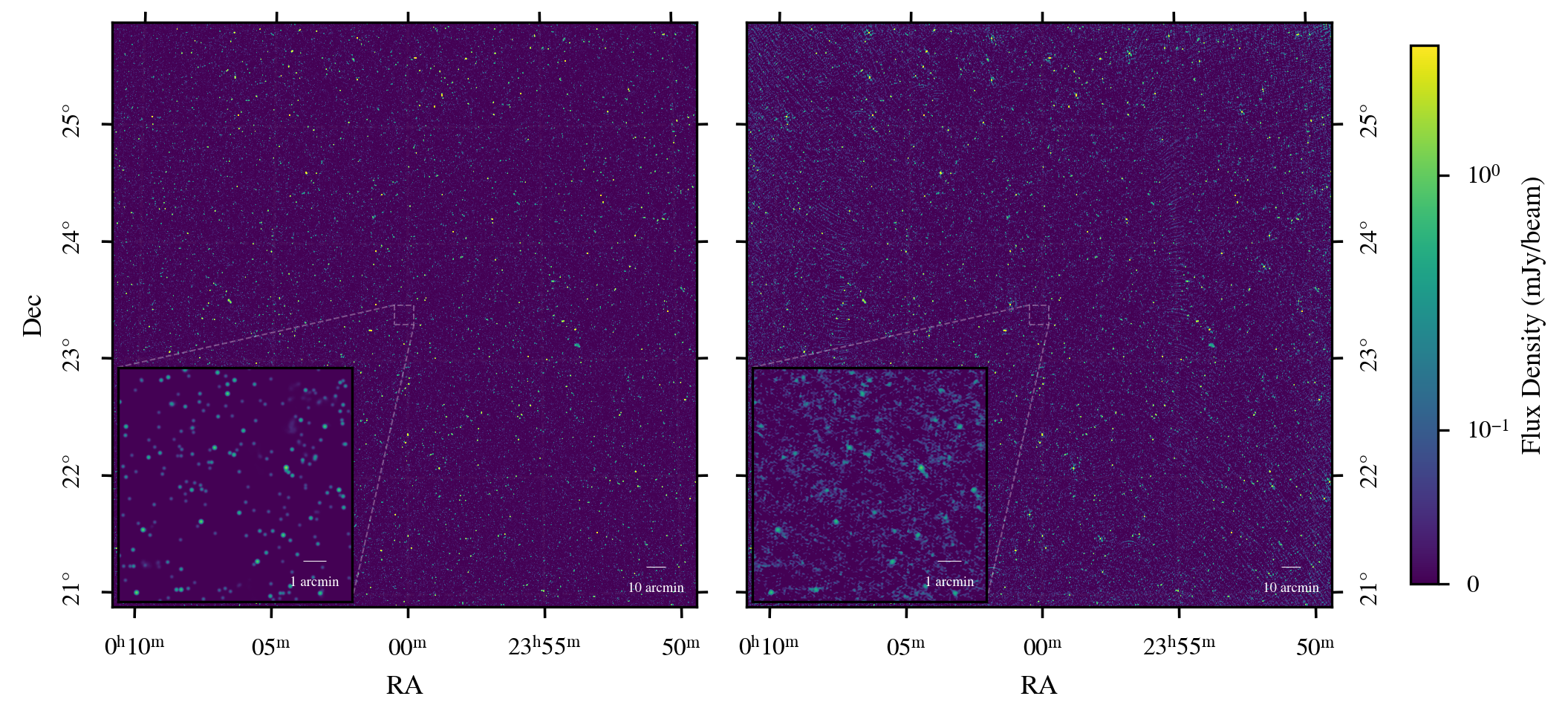}
    \caption{Image of the simulated sky model (left) and the corresponding sky map (right). For both, the inlet shows a \ang{;10} center cutout. Colors are scaled and limited for visualization purposes.}
    \label{fig:map-5deg-cutout}
\end{figure*}

\begin{figure*}
    \centering
    \includegraphics[width=17cm, trim=0 2mm 0 0, clip]{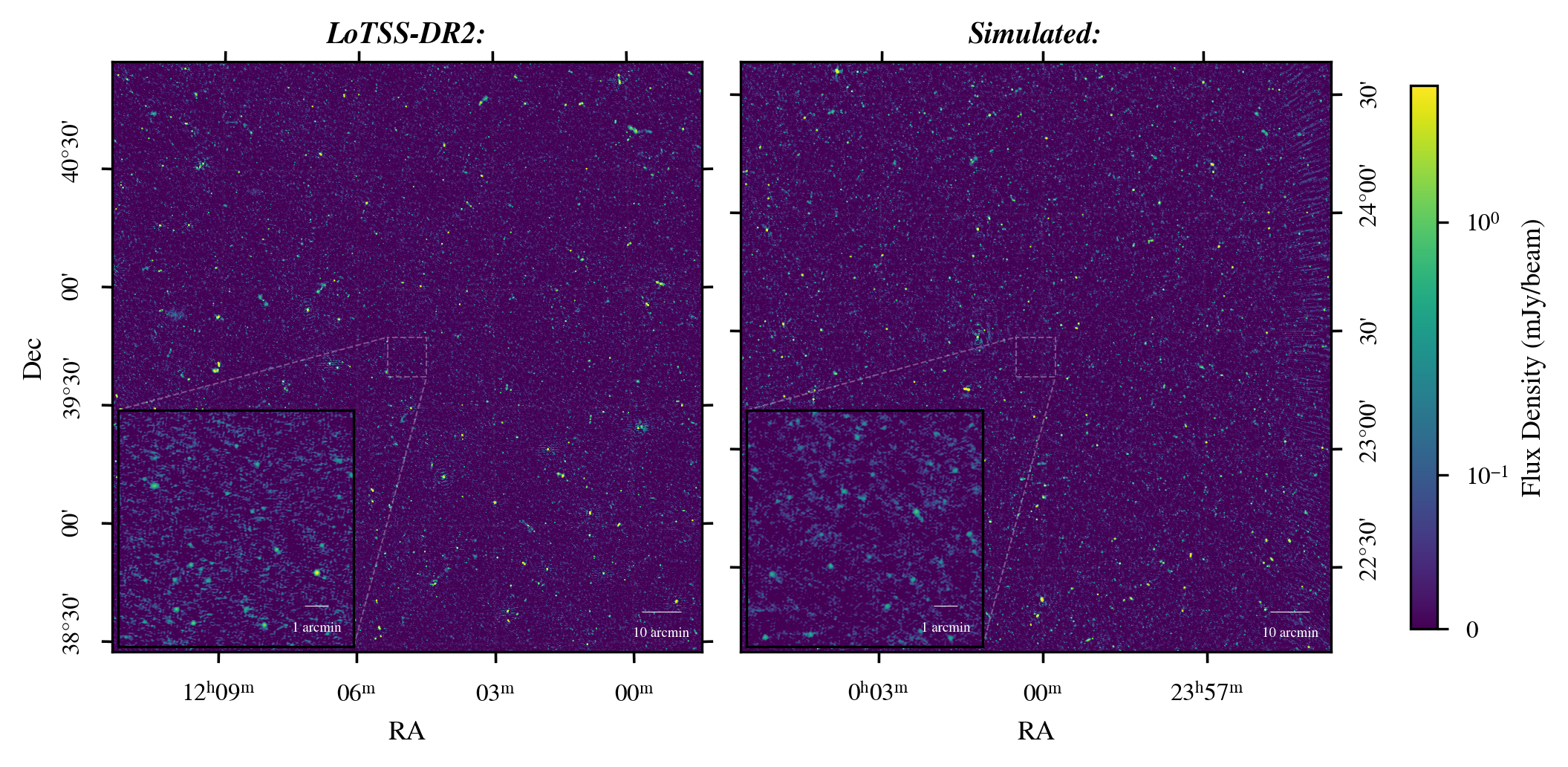}
    \caption{Side-by-side display of a real LoTSS map (left) and our simulated map (right). The LoTSS image is  from the P181+40 mosaic \citep{lotss-dr2:P181-40}. In both cases, the image is a \ang{2.5} center cutout and the inlet shows a \ang{;10} center cutout.}
    \label{fig:map-vs-lotss-plot}
\end{figure*}

To evaluate our simulation, we generated a \ang{5}$\times$\ang{5} sky map as described in Section~\ref{sec:simulation-of-radio-survey-maps}. The numbers of sources sampled with the DM and modeled as 2D-Gaussians are \num{12749} and \num{196759}, respectively. In Fig.~\ref{fig:map-5deg-cutout}, we show the entire sky model and sky map side by side to provide an impression of the resulting products. We converted the sky map from \si{\Jy \per \px} to \si{\Jy \per \beam} to have both maps in the same unit. The shared color scale is optimized to visualize the large range of flux values present in both maps. Artifacts in the sky map, such as striations or spots around bright sources, are a typical side-effect of the cleaning procedure caused by bright sources in the FOV. In addition, we offer a side-by-side comparison of a LoTSS pointing and our simulated map in Fig.~\ref{fig:map-vs-lotss-plot}. The LoTSS image was taken from the mosaic of field P181+40 \citep{lotss-dr2:P181-40}, and both images are a \ang{2.5} center-cutout of the respective map.\\

\begin{figure}
    \centering
    \begin{subfigure}[t]{\hsize}
        \centering
        \resizebox{\hsize}{!}{\includegraphics{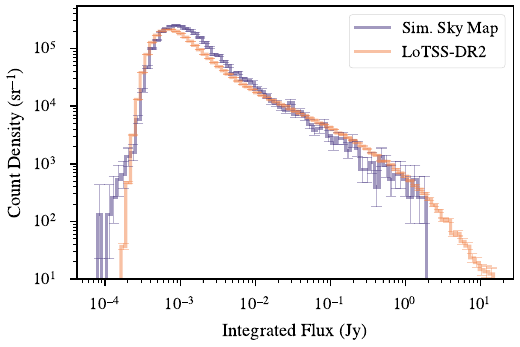}}
        \label{fig:sub:source-catalogs-int-flux}
    \end{subfigure}
    \begin{subfigure}[t]{\hsize}
        \centering
        \resizebox{\hsize}{!}{\includegraphics{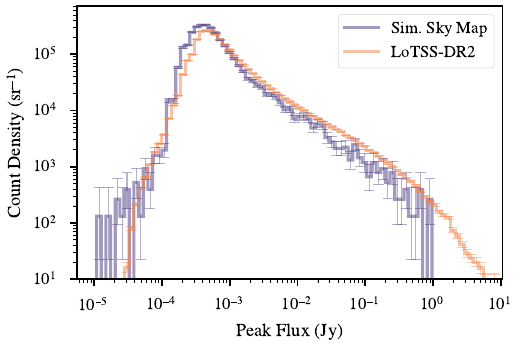}}
        \label{fig:sub:source-catalogs-peak-flux}
    \end{subfigure}
    \begin{subfigure}[t]{\hsize}
        \centering
        \resizebox{\hsize}{!}{\includegraphics{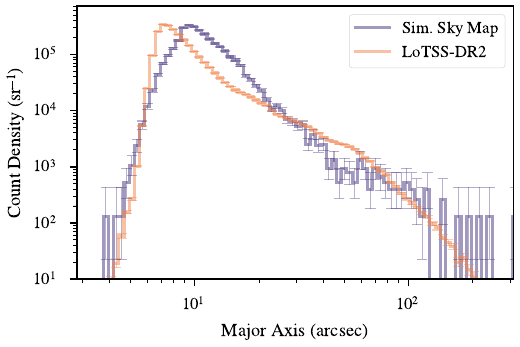}}
        \label{fig:sub:source-catalogs-major-axis}
    \end{subfigure}
    \caption{Histograms of the integrated flux, peak flux, and major axis values, showing the distributions over sources in the catalogs of the simulated sky map the LoTSS-DR2 maps.}
    \label{fig:source-catalogs-histograms}
\end{figure}

In order to compare the real and simulated maps as laid out in Section~\ref{sec:sub:properties-of-the-simulated-sources}, we ran \textsc{PyBDSF} on the simulated map to obtain the map catalog. To align with the work of \cite{Shimwell+22}, we used the settings shown in Table~\ref{tab:pybdsf-settings}, all other values were left to their default. Based on this catalog and the LoTSS-DR2 catalog, we computed the histograms for the aforementioned integrated flux, peak flux and major axis of the modeled sources.
To normalize the histograms, we divided the counts by the total map area, i.e., histograms are given in units of \si{\per\steradian}. For the simulated maps, this area corresponds to the map size of $\SI{25}{\deg\squared} \approx \SI{7.615e-3}{\steradian}$, whereas for the LoTSS data, this corresponds to the published survey area \citep{Shimwell+22} of $\SI{5634}{\deg\squared} \approx \SI{1.716}{\steradian}$. The histograms are shown in Fig.~\ref{fig:source-catalogs-histograms}, error margins again corresponding to the $1\sigma$ frequentist confidence intervals.\\
Regarding the integrated and peak fluxes in the top and middle plots, the sky map distributions match the LoTSS distributions fairly well. The integrated flux distribution of the simulated map is cut off at \SI{1}{\Jy}, which is a consequence of the exclusion of bright sources in the simulation. Apart from that, the maximum of the distribution is slightly shifted toward higher values, where a small excess of sources in the range of 1 to \SI{10}{\milli\Jy} as compared to the LoTSS data is observed. Further experimentation reveals that this excess is attributed to limitations in the cleaning procedure. Sources close to the cleaning threshold of 0.5 times the RMS are not fully deconvolved, resulting in an increase of both the integrated flux density and the size for these sources. This effect becomes increasingly relevant for smaller source fluxes. We observe that the shift in the distribution can be reduced through deeper cleaning, for example by adapting the stopping criteria for the cleaning algorithm or by running the simulation with a lowered threshold for exclusion of bright sources, whose presence induces artifacts which limit the cleaning depth. The same experiments show that the shift is correlated with the residual RMS value discussed below in Section~\ref{sec:sub:residual-rms}. This overestimation of source signal counteracts the expected lowering of the peak flux values discussed in Section~\ref{sec:sub:properties-of-the-simulated-sources}, which is nonetheless still observed in the middle plot as a small shift toward smaller values in the simulated distribution. The real LoTSS observations are less affected by these issues due to differences in the imaging procedure, which are discussed in Section~\ref{sec:discussion}.\\
Regarding the sizes in the bottom plot, the LoTSS distribution peaks just above the restoring beam size of $\sim$\ang{;;6}, and the simulated distribution just above $\sim$\ang{;;8.5}, which is in accordance with the expected effective resolution. There seems to be a scarcity of larger sources in the simulation, covering the range of \ang{;;20} to \ang{;;70}. This regime is populated by extended sources, whose sizes are sampled from the distribution found in the training data. Therefore, this discrepancy could be an indicator of a bias in the data selection procedure that disfavors larger sources. This possibility is discussed in Section~\ref{sec:discussion}. Nonetheless, the simulated distribution approximately resembles the LoTSS distribution in its shape, as well as the overall covered size range.\\
\begin{table}
    \caption{Settings applied in \textsc{PyBDSF} for the analysis of our simulated maps.}
    \label{tab:pybdsf-settings}
    \centering
    \begin{tabular}{l l}
        \hline\hline
        Setting & Value \\
        \hline
        atrous\_do & True \\
        thresh\_isl & 4 \\
        thresh\_pix & 5 \\
        adaptive\_rms\_box & True \\
        beam & (\num{1.667e-3}, \num{1.667e-3}, 0) \\
        \hline
    \end{tabular}
\end{table}

%-----------------------------------------------
\subsection{Residual RMS}
\label{sec:sub:residual-rms}

\begin{figure}
    \centering
    \resizebox{\hsize}{!}{\includegraphics[trim= 0 10mm 0 10mm, clip]{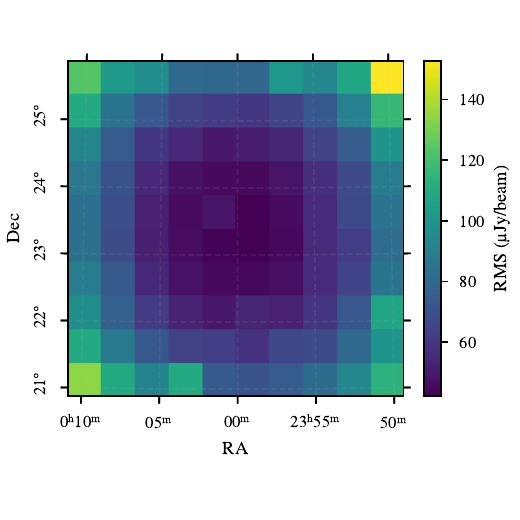}}
    \caption{RMS values of the \textsc{DDFacet} residual image, individually calculated for 100 different square tiles of the entire image.}
    \label{fig:residual-rms}
\end{figure}

With the simulated map, we proceeded to calculate the RMS value of the residual image that is output by \textsc{DDFacet}. We further divided this residual image into a grid of \numproduct{10 x 10} square-shaped tiles, each of size \ang{0.5}$\times$\ang{0.5}, and individually calculated the RMS within each of those tiles. The result is shown in Fig.~\ref{fig:residual-rms}.\\
The residual map has an overall RMS value of \SI{76.3}{\micro\Jy\per \beam} and is therefore compatible with the median RMS sensitivity on the LoTSS-DR2 maps of \SI{83}{\micro\Jy\per \beam} reported in \cite{Shimwell+22}. The RMS for individual tiles varies across the FOV between a minimum and maximum of \SI{42.2}{\micro\Jy\per \beam} and \SI{153}{\micro\Jy\per \beam}, respectively, which corresponds to a maximum variation by a factor of $\sim$\num{3.6}. As expected, the RMS across the FOV qualitatively resembles the Gaussian-shaped LOFAR primary beam profile, with higher RMS at increasing distance from the image center. Occasional deviations from this profile can be attributed to the presence of bright sources within the respective tiles. 

%%%%%%%%%%%%%%%%%%%%%%%%%%%%%%%%%%%%%%%%%%%%%%%%
\section{Discussion} 
\label{sec:discussion}

In this work we provided a method for simulating realistic LOFAR telescope survey observations, utilizing a generative machine learning model to synthesize radio galaxy images with precise morphological details authentic to real sources. Moreover, our multi-step approach provides realism across different properties of individual sources regarding positions, fluxes, and sizes. It also provides an accurate representation of effects caused by the time and frequency coverage, the incomplete \textit{uv}-coverage, the station layout, and the model primary beam. Concurrently, it allows  independent control over these properties, as each can be manually adapted to individual needs. For instance, the source positions can be randomized independently from other properties, making this software potentially useful for studies of statistical cosmology, for example \cite{Hale+23}. Similarly, the incorporation of \textsc{LoSiTo} allows   the inclusion of additional instrumental effects apart from noise, including ionospheric and station-based systematic effects \citep{Edler+21}. Moreover, our simulation not only provides a sky map that recreates the result of a LoTSS-like LOFAR pointing as reconstructed with the current standard analysis methods, but also the corresponding sky model. This further expands the realm of possible application scenarios. For instance, developers can test new methods of data analysis for tasks such as calibration or imaging on the simulated observations and can compare their result both to the sky model and the standard analysis.\\
Our simulation can generate realistic synthetic radio images that reflect a broad range of source populations, morphologies, and instrumental effects. This is useful, for example, for validating detection pipelines, as is attempted, for example with the SKA data challenges \citep{2021MNRAS.500.3821B}, or for testing automatic source detection software, such as the Selavy pipeline \citep{2023PASA...40...27B}.
The software can also be used to run end-to-end simulations testing the pipeline, from visibility to final catalogs, and to augment existing training data for machine-learning aided detection algorithms or to test self-supervised machine learning models \citep{2025arXiv250319111B}. Another use case comes from the use of large-scale radio models to constrain cosmological parameters \citep{2025arXiv250420723P}. These models require comparison of survey statistics against mock catalogs that include incompleteness and measurement errors, accounting for sensitivity variations across the sky and their impact on source detection completeness and properties.\\
Our improved selection and pre-processing of training data has brought significant improvement to the quality of the images sampled by the DM. Compared to the results presented in \cite{VicanekMartinez24}, pixel statistics are now reproduced in greater detail. We believe this improvement is likely related to the use of source masks, which not only allows the DM to focus on relevant image details through the rejection of background signals, but also facilitates the use of random rotations of arbitrary angles as a means of data augmentation during training, which significantly extends the data manifold.\\
Nonetheless, it is important to carefully consider biasing effects resulting from the selection procedure, some of which are possible to recognize, while others might be challenging to identify in practice. As we laid out in Section~\ref{sec:sub:training-data}, we excluded over \SI{80}{\percent} of the training data, where the largest chunk is due to high pixel values on the source mask edges. This choice represents a clear preference of sample purity over completeness, where it is possible to relax the exclusion criterion and include a larger fraction of possibly useful images, which results in a denser sampling of the data manifold, albeit at the expense of possibly keeping examples with unrealistic morphological features caused by erroneous segmentation of the original cutout. Moreover, our sigma-thresholding-based masking procedure can lead to incomplete coverage of multi-component sources. For instance, we observed through visual inspection that in some instances of double-lobed sources with high asymmetry in brightness, only one of the two jets was masked, i.e., one-half of the source was omitted. While this did not necessarily lead to unrealistic training examples, it did induce a bias in the training distribution, for example an over-representation of single-lobed sources in the case of the mentioned example. A similar argument can be made for the reverse case, where distinct close-by sources might be covered by the same mask, resulting in training images that contain more than one source. In addition, by discarding all examples where the source mask had more than one single island, we systematically removed all sources with two lobes that are not connected by jet emission.\\
A possible way to improve the determination of source masks and potentially mitigate those issues is through the use of source models. The \textsc{PyBDSF} software creates models of source morphologies as combinations of elliptical 2D Gaussians, which can be reconstructed from the publicly available component catalog \citep{Hardcastle+23}. These models could be used to construct masks in a more accurate way, without having to resort to sigma thresholding. The catalog was curated through multiple iterations of visual inspection by  citizen scientists and by experts, providing the source characterizations with an increased reliability that the training data could directly benefit from. Another advantage is that every mask could be exactly associated with the corresponding source, even in cases of multi-component sources where the masks have multiple disconnected islands. Further, this would also introduce new possible selection criteria for source visibility, such as S/N estimates based on statistical errors of the fitted flux density values. This could again potentially facilitate a more accurate and less discriminatory method for data selection.\\
At the current stage, another limitation of the DM is the size limit, as the model was trained on images of \SI{80}{\px} side length, which with the LoTSS-DR2 pixel resolution corresponds to \ang{;;120}. Currently, sources larger than this are generated by rescaling smaller images. This is an inaccurate way of modeling large extended source morphologies, as the nominal resolution of the image is effectively decreased through the up-scaling. Put in simple terms, a point-like feature will not look like the restoring beam, but instead will be extended proportionally to the up-scaling factor. This can be remedied by increasing the size of the DM, which reduces or even eliminates the need for up-scaling. The relevance of this issue will be further increased with next-generation radio telescopes operating at higher resolutions. To increase computation efficiency, separate models can be employed for different size scales.\\
A limitation that is fundamental to our approach is the lower effective resolution of \ang{;;8.5}, as compared to the \ang{;;6} resolution sky maps, the reason for which is explained in Section~\ref{sec:sub:simulated-observations-and-imaging}. The only way to circumvent this issue is by training the DM on higher-resolution data, which would in turn allow   the generation of a higher-resolution sky model. One potential method would be to construct training images from hydrodynamical simulations of AGN, such as those in \cite{YatesJones+23}. While solving the mentioned issue, this might reduce the morphological variability of radio sources found in real observations, depending on what morphologies can be reproduced with the employed simulation.\\
As laid out in Section~\ref{sec:sub:source-fluxes-and-sizes}, the characteristics of the simulated maps are sensitive to the depth of the cleaning, where incomplete deconvolution leads to distortions in the frequency distribution of low-flux sources. This issue is somewhat mitigated by excluding very bright sources, which cause brighter artifacts in the deconvolved image. This in turn adds a hard cut in the upper range of the flux distribution that is not representative of real observations. There is a limit to how closely the real observations can be matched in this regard, as the cleaning procedure used in the simulation is different to the one employed for the real LoTSS maps. For those, a sophisticated pipeline of many alternating calibration and imaging steps was implemented, also including different refinements of the flux density scale, see \cite{Shimwell+22} and references therein. In addition, our maps simulate a single pointing, whereas the real maps are combined through the mosaicing of several overlapping pointings, which additionally reduces effects such as the uneven noise level due to the primary beam response and the time- and frequency-smearing of sources farther away from the phase center.\\
The core assumption that the integrated flux density of the source is not related to its morphological features may not be true. For instance, \cite{Saripalli+12} report such a relation, based on a limited sample of radio sources obtained with the Australia Telescope Low Brightness Survey at \SI{1.4}{\giga\hertz}. Using the standard Fanaroff-Riley (FR) classification scheme \citep{Fanaroff&Riley1974}, they concluded that the fraction of observed FR type-I morphologies increases with decreasing flux. These effects could in principle be accounted for by controlling source morphologies during sampling, a possibility demonstrated in \cite{VicanekMartinez24}. This would also allow   a more realistic representation of SFGs, where our simulation currently employs a simplified morphological model of elliptical 2D Gaussians. On the other hand, the training data contains not only images of AGN, but also SFGs. There are no exact numbers on the proportions, as reliable AGN selection is still an ongoing effort \citep{Hardcastle+23}. This circumstance is likely to produce a bias toward higher counts of SFGs in our simulations. All these issues can be addressed through class- or morphology-controlled sampling, for which a training dataset with high-quality morphological labels derived from a versatile classification scheme and reliable selection criteria would be required. A possible groundwork toward the generation of such a dataset is given in \cite{Drake+2024}, where the authors combine the use of Monte Carlo simulations and spectral analysis to determine the reliability of galaxies belonging to each of four classes of objects (SFGs, radio-quiet AGN, and high- and low-excitation radio galaxies) for a subset of the LoTSS-DR2 catalog. By adding the use of those tags to our method of data selection and preparation, a labeled image dataset can be created to train a DM for targeted sampling. Setting a threshold on the given reliability would allow   control over sample purity versus completeness.\\
Finally, the fact that sizes of extended sources are sampled from the training data distribution translates a possible selection bias directly into the simulated observations. For instance, the bias toward single-lobe sources implies a bias toward smaller extensions, which is a possible explanation for the scarcity of sources at \ang{;;20} to \ang{;;70} observed in the simulated maps. This can be remedied in conjunction with the improved masking procedure proposed earlier in this section, where one could use values from the source catalog to parameterize the size, rather then relying on the source masks. This way, sizes can directly be sampled from the observed distribution, making this step independent from biases in the training data. Another limitation related to sizes is that our simulation cuts off at a few arcminutes, disregarding the presence of radio galaxies with extremely large angular sizes on the order of \ang{;10}. While these galaxies could technically be sampled with models of larger image size, it is questionable whether there are enough available training images of such objects.  \cite{Hardcastle+23} report that the optical cross-match catalog contains only 89 sources with sizes larger than \ang{;10}. It is open to question whether such numbers might be enough for possibly fine-tuning a pre-trained model or whether a different approach is required for this.\\
%

%%%%%%%%%%%%%%%%%%%%%%%%%%%%%%%%%%%%%%%%%%%%%%%%
\section{Data availability}
The code implemented in this work will be made available at \url{https://github.com/tmartinezML/MASSIMO}. Finished products of simulations can be shared on request.\\
\begin{acknowledgements}
We thank the referee for a constructive report. MB and TVM acknowledge support by the Deutsche Forschungsgemeinschaft under Germany’s Excellence Strategy – EXC 2121 Quantum Universe – 390833306 and via the KISS consortium (05D23GU4) funded by the German Federal
Ministry of Education and Research BMBF in the ErUM-Data action plan.\\
\end{acknowledgements}

\bibliographystyle{aa}
\bibliography{aa54794-25corr}

\begin{appendix}

%%%%%%%%%%%%%%%%%%%%%%%%%%%%%%%%%%%%%%%%%%%%%%%%
\section{Determination of source masks}
\label{app:determination-of-source-masks}
\begin{figure}[h!]
    \centering
    \resizebox{\hsize}{!}{\includegraphics{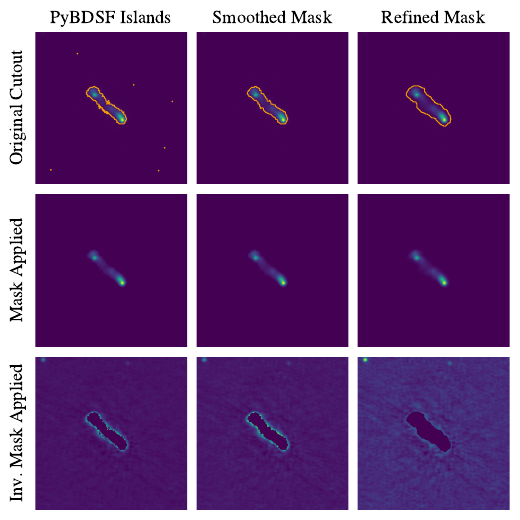}}
    \caption{Depiction of post-processing steps applied to the source mask of a random example from cutouts selected as training data.}
    \label{fig:mask-post-processing}
\end{figure}

\textsc{PyBDSF} \citep{pyBDSF} is a software commonly used for radio source detection and modeling. As part of the source detection algorithm, it identifies islands of contiguous source emission by separating between source and noise pixels based on a threshold value derived from pixel statistics. For more details, we refer to the documentation.\footnote{\url{https://pybdsf.readthedocs.io/en/latest/}}
We obtained those islands by running \textsc{PyBDSF} on all images individually, using an island threshold of 3.5, a pixel threshold of 1, and default values for all other settings. These islands were used as a base for our source masks. We further smoothed those masks by first filling any holes in the islands, and subsequently performing a binary closing operation with a $3\times3$ kernel followed by a binary opening operation with a $1\times1$ kernel. Finally, we removed all islands smaller than eight pixels in size.\\
We then proceeded to apply a final refinement to the source masks by extending their boundaries based on the local noise level. For this, we defined a local background region constrained by an ellipse around the source mask centroid, such that this background region covers at least 20 times the number of pixels contained in the source mask, or the entire image background region, depending on which is smaller. For this local background region, we calculated the median and the standard deviation. The source mask was then expanded such that it covers all neighboring pixels that have a value above the median plus five times the standard deviation. This entire procedure was performed iteratively until the mask size converged, but at most for 20 iterations. We find the combination of island smoothing and expansion to generally improve the coverage of source signal on the images, which is exemplary showed in Fig.~\ref{fig:mask-post-processing}.\\
For the generated samples, source masks were determined in a similar way. Instead of starting with the \textsc{PyBDSF} mask, we first calculated the mean and standard deviation of all pixels outside the inner circle of the image. Since the training images are prepared in a way that no source signal lies outside this inner circle, i.e., this area is zero for all pixels in all the training images, we expect only residual noise from the diffusion sampling process in this part of the sampled images. We then determined our initial source mask as those pixels on the image that have values larger than this mean plus three times the standard deviation. Following that, we applied the smoothing procedure as described above, and used the result as the sample source mask.

\FloatBarrier
\section{Example cutouts from the image selection procedure}
\label{app:example-cutouts}
\begin{figure}
    \centering
    \resizebox{\hsize}{!}{\includegraphics{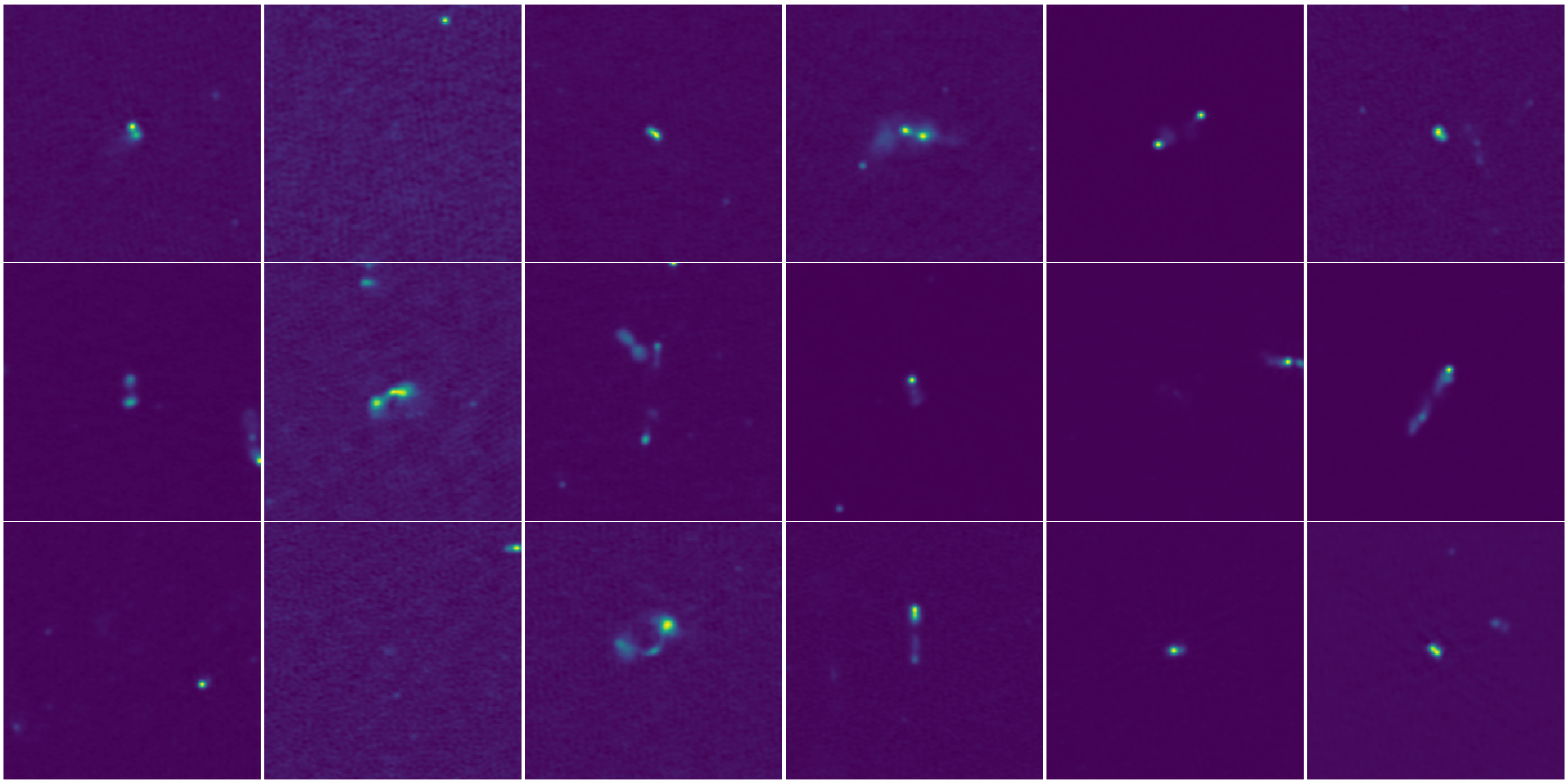}}
    \caption{Random examples of cutouts from the initial selection after applying the $S/N_\sigma$ cut.}
    \label{fig:example-grid-initial-selection}
\end{figure}

\begin{figure}
    \centering
    \resizebox{\hsize}{!}{\includegraphics{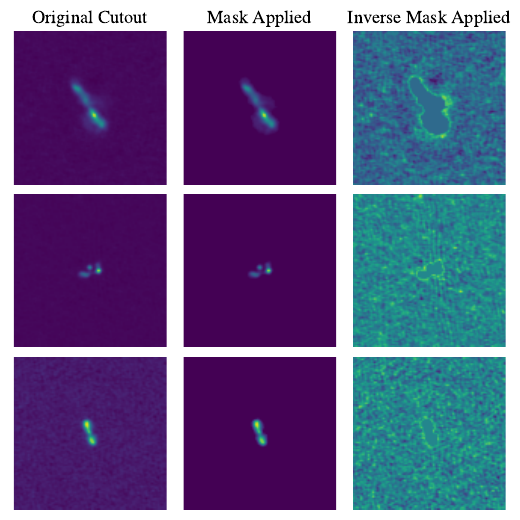}}
    \caption{Random examples of cutouts excluded due to large pixel values on the mask edges.}
    \label{fig:mask-edge-threshold-cutouts}
\end{figure}

We show a few examples of cutouts as obtained or excluded at different stages of the data selection procedure described in Section~\ref{sec:sub:training-data}.\\
In Fig.~\ref{fig:example-grid-initial-selection}, we show a selection of images from the cutouts as obtained by cropping around the positions in the optical cross-match catalog. Some sources are clearly visible, while other cutouts contain high levels of noise. Those are excluded in the first step of applying the $S/N_\sigma$-cut, as is shown in Fig.~\ref{fig:example-grid-initial-selection}. In Fig.~\ref{fig:mask-edge-threshold-cutouts}, we show three images that are excluded due to large pixel values on the mask edge. The top and bottom row show examples where the masking leads to unrealistically hard edges. This is however less obvious in the example of the center row, which demonstrates how this choice represents a trade-off between sample completeness and data quality.

\begin{figure}
    \centering
    \resizebox{\hsize}{!}{\includegraphics{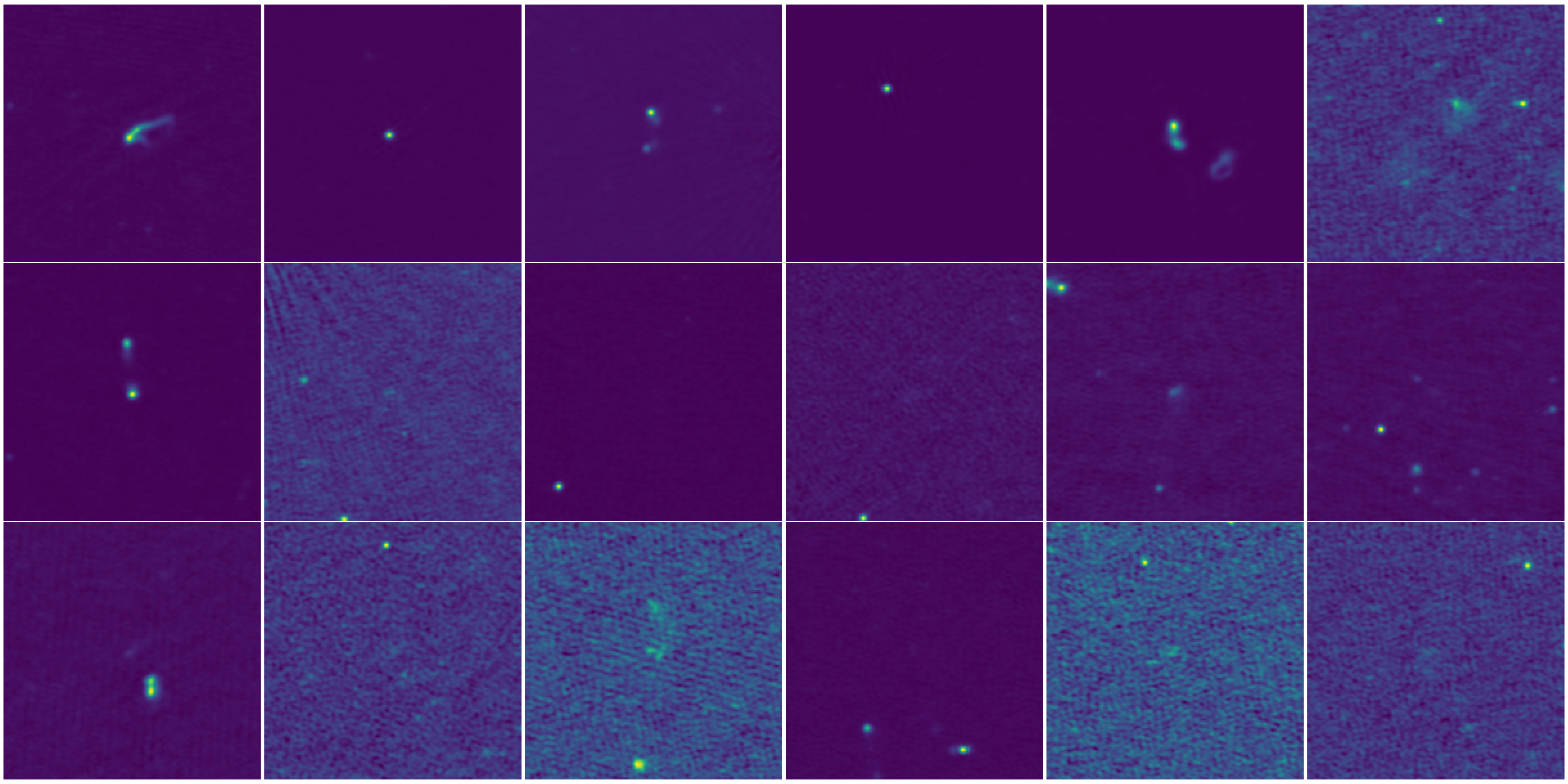}}
    \caption{Random examples of all cutouts as obtained from the optical cross-match catalog.}
    \label{fig:example-grid-all-cutouts}
\end{figure}

\FloatBarrier
\section{Details on model architecture and training}
\label{app:details-on-model-architecture-and-training}

\begin{figure}[h!]
    \centering
    \resizebox{\hsize}{!}{\includegraphics{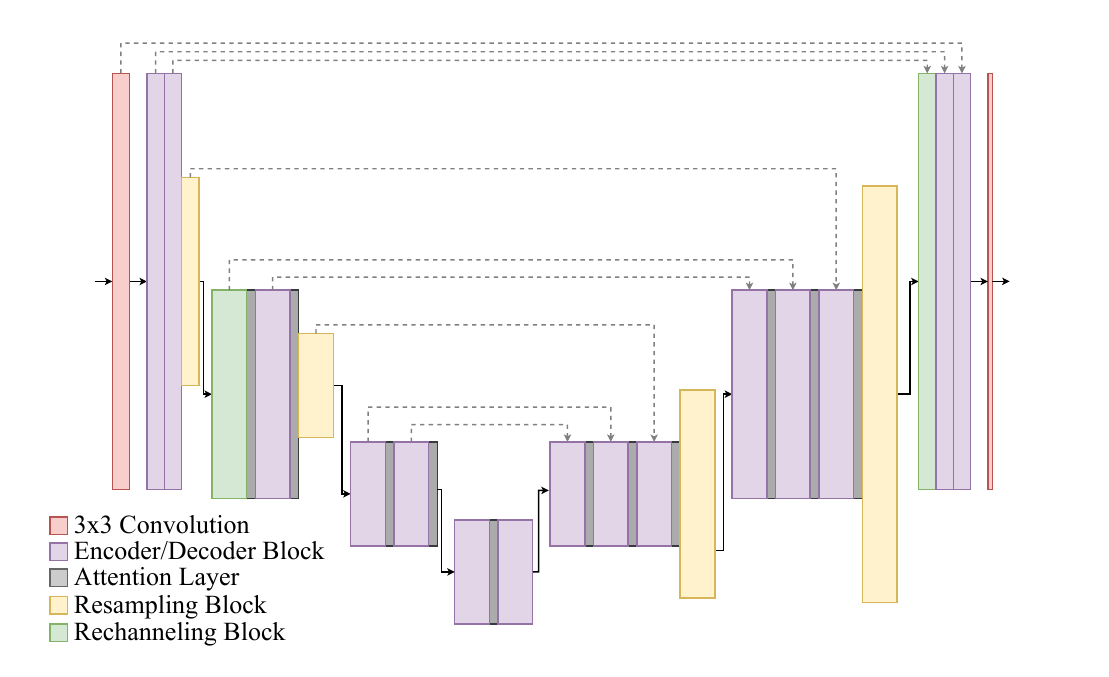}}
    \caption{Architecture of the U-Net. Skip connections between encoder and decoder are indicated with dashed lines.}
    \label{fig:unet}
\end{figure}

\begin{figure}[h!]
    \centering
    \resizebox{\hsize}{!}{\includegraphics{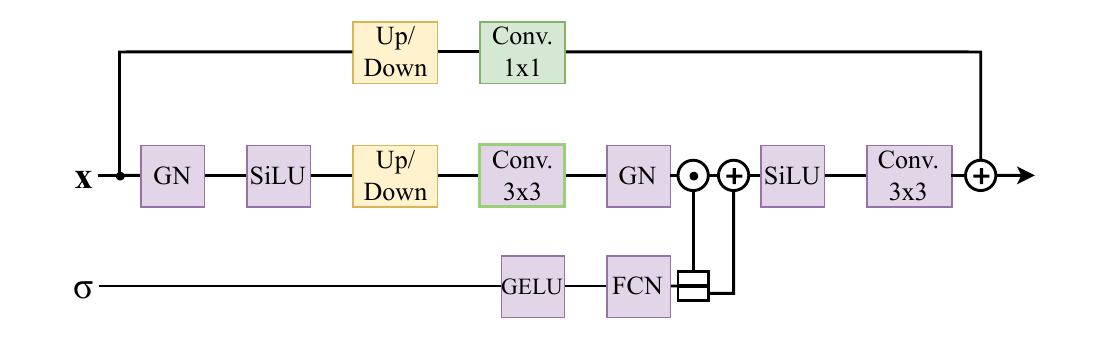}}
    \caption{Architecture of the U-Net building block, including group normalization (GN), sigmoid linear unit (SiLU), and convolutional layers. Conditioning information is embedded through a Gaussian error linear unit (GELU) layer and a subsequent fully connected network (FCN). Variable elements are colored as in Fig. \ref{fig:unet}. The first $3\!\times\!3$ convolution is always included; only the change in channel dimensions is optional. }
    \label{fig:unet_block}
\end{figure}

In the following, we offer a description of the employed model architecture and training procedure. For a more detailed explanation, we refer to \cite{VicanekMartinez24}.\\
We used a U-Net architecture \citep{Ronneberger+15} with convolutional neural network (CNN) blocks for our denoiser network, illustrated in Fig. \ref{fig:unet}. It consists of an encoder part, a bottleneck part and a decoder part. Both encoder and decoder hold three levels of different resolution and channel depth, with corresponding resampling and rechanneling operations in between. Levels of equal resolutions in the encoder and decoder are linked through skip connections.
All elements of the network are constructed from the same block, shown in Fig. \ref{fig:unet_block}, with slight variations depending on the function in different parts of the network. These variations are illustrated with colors, consistent between Figures \ref{fig:unet} and \ref{fig:unet_block}. In the lower-resolution levels and the bottleneck part, additional self-attention layers are added. Architectural details are given in Table \ref{tab:U-Net}.\\
Since the network is trained to remove noise from training images, we employed the $L_2$ loss between the noisy image and the denoiser output. This reads
\begin{equation}
    \mathcal{L}(\theta) = c_{\mathrm{out }}(\sigma)^{-2} \left\Vert D_\theta(\mathbf{x}+\mathbf{n} ; \sigma)-\mathbf{x}\right\Vert_2^2,
\end{equation}
where $\theta$ are the parameters of the denoiser $D_\theta$, $\mathbf{x}$ is the training image, and $\mathbf{n}$ is the added Gaussian noise with standard deviation $\sigma$, i.e., $\mathbf{n} \sim \mathcal{N}\left(\mathbf{0}, \sigma^2 \mathcal{I}\right)$. We note that $\sigma$ is varied throughout the training. The factor $c_{\mathrm{out }}(\sigma)$ is added to balance the loss magnitude across different values of $\sigma$. A detailed definition and more details on the variation of $\sigma$ are given in \cite{VicanekMartinez24}.\\
We trained the model for \num{150000} iterations with a batch size of 256. We used dropout as well as context dropout, both with a rate of 0.1. We kept an exponential moving average of the model weights during training, with a rate of 0.9999.

\begin{table}[h!]
    \caption{Details of the implemented U-Net architecture.}
    \label{tab:U-Net}
    \centering
    \begin{tabular}{l l}
        \hline\hline  
        Parameter & Value \\
        \hline
        Resolution levels & $(80, 40, 20)$ \\
        Initial channels & \num{128} \\
        Channel multipliers & $(1, 2, 2) $ \\
        Norm. groups & \num{32} \\
        Attention & (No, Yes, Yes) \\
        Attention heads & \num{2} \\
        Attention head channels & \num{32} \\
        \hline
    \end{tabular}
    \tablefoot{
         Resolution levels indicate the image sizes in pixels at the different levels in the U-Net. The number of channels at different levels is given as a product of the initial channels and corresponding channel multiplier.
    }
\end{table}

\FloatBarrier
\section{Technical details of simulated observations and imaging}
\label{app:technical-details-of-visivilities-and-imaging}
We used \textsc{LoSiTo}'s synthms script to simulate empty LOFAR HBA MSs over the entire HBA frequency range, with an observation time of \SI{8}{\hour}, a time resolution of \SI{8}{\second}, and two channels per sub-band. For the starting time and observation position, we adopted values corresponding to the P000-23 pointing \citep{lotss-dr2:P000-23}, with right ascension at $00^{\mathrm{h}}00^{\mathrm{m}}07.5^{ \mathrm{s}}$, declination at  $+23^\circ23{}^\prime42.9036{}^{\prime\prime}$ and the start of the observation happening at 2018-11-30 15:14:39.006. This results in 246 MSs, each representing one sub-band with two channels. We then applied the DP3 average step to combine those into 24 MSs of 20 sub-bands each, whereby the first as well as the last five of the initial MSs were excluded. In practice, this step is done only once to create a template of MSs. For a new map simulation, those files are copied, and the copies are used for further processing.\\
Cleaning was done using the Högbom algorithm in \textsc{DDFacet}, with a restoring beam size of \ang{;;6} and a pixel size of \ang{;;1.5}, and using the LOFAR station beam model. Parameter settings for deconvolution are summarized in Table~\ref{tab:ddfacet-settings}. We used a cleaning mask, which was constructed selecting all pixels in the sky model above a threshold value that is set to \SI{5e-5}{\Jy \per \beam}
\begin{table}[h!]
    \caption{Settings applied in \textsc{DDFacet} for cleaning.}
    \label{tab:ddfacet-settings}
    \centering
    \begin{tabular}{l l}
        \hline\hline
        Setting & Value \\
        \hline
        MaxMajorIter & \num{20} \\
        MaxMinorIter & \num{2e7} \\
        RMSFactor & \num{0.5} \\
        PeakFactor & \num{0.2} \\
        \hline
    \end{tabular}
    \tablefoot{We used the Högbom-\textsc{clean} deconvolution algorithm on our simulated maps. All other parameters were left to their default values.}
\end{table}

\end{appendix}

\end{document}